\documentclass[epsf,twocolumn,showpacs,preprintnumbers]{revtex4}
\usepackage{graphics}
\usepackage{graphicx}
\usepackage{dcolumn} 
\usepackage{bm}
\usepackage{epsfig}
\begin{document}
\title{Infinite Layer LaNiO$_2$: 
  Ni$^{1+}$ is not Cu$^{2+}$}
\author{K.-W. Lee and W. E. Pickett}
\affiliation{Department of Physics, University of California, Davis, CA 95616}
\date{\today}
\begin{abstract}
The Ni ion in LaNiO$_2$ has the same formal ionic configuration $3d^9$ as
does Cu in isostructural CaCuO$_2$, but it is reported to be nonmagnetic
and probably metallic whereas CaCuO$_2$ is a magnetic insulator.  From
{\it ab initio} calculations we trace its individualistic behavior to 
(1) reduced $3d-2p$ mixing due to an increase of the separation of
site energies ($\varepsilon_d - \varepsilon_p$) of at least 2 eV, and
(2) important Ni $3d(3z^2-r^2)$ mixing with La $5d(3z^2-r^2)$ states that leads to
Fermi surface pockets of La $5d$ character that hole-dope the Ni $3d$
band.  Correlation effects do not
appear to be large in LaNiO$_2$.  However, {\it ad hoc} increase of the 
intraatomic repulsion on the Ni site (using the LDA+U method)
is found to lead to a novel correlated state:
(i) the transition metal $d(x^2-y^2)$ and $d(3z^2-r^2)$ 
states undergo consecutive Mott
transitions, (ii) their moments are {\it antialigned}  
leading (ideally) to a ``singlet" ion in which there are
two polarized orbitals, and (iii) 
mixing of the upper Hubbard $3d(3z^2-r^2)$ band with the La $5d(xy)$ states
leaves considerable transition metal $3d$ character in a band 
pinned to the Fermi level.
The magnetic configuration is more indicative of a Ni$^{2+}$ ion in this
limit, although the actual charge changes little with $U$.
\end{abstract}
\maketitle

\section{Introduction}
The perovskite oxide LaNiO$_3$, purportedly an example of a correlated metallic 
Ni$^{3+}$ system, has been 
investigated over some decades by a few 
groups\cite{goodenough,sreedhar,gayathri} for possible exotic behavior.
The oxygen-poor lanthanum nickelate LaNiO$_x$ 
has also attracted attention,
because of characteristic changes of its electronic and magnetic properties 
as the oxygens are removed.
It is metallic at $2.75 < x < 3$, but semiconducting 
for $2.50 < x < 2.65$.\cite{moriga}
For $x=2.6$, it shows ferromagnetic ordering with 1.7 $\mu_B$/Ni
below 230 K \cite{moriga}
and magnetic behavior of the $x=2.7$ material
has been interpreted in terms of a model of 
ferromagnetic clusters.\cite{okajima}
At $x=2.5$, where formally the Ni is divalent,
a perovskite-type compound La$_2$Ni$_2$O$_5$ forms in which 
NiO$_6$ octahedra lie along  $\it{c}$ axis directed chains
and NiO$_4$ square-planar units alternate in the
$\it{a-b}$ plane.  This compound shows 
antiferromagnetic ordering of the NiO$_6$ units along the $\it{c}$ axis 
but no magnetic ordering of the NiO$_4$ units.\cite{alonso}

Since LaNiO$_2$ with formally monovalent Ni ions was synthesized 
by Crespin {\it et al.}\cite{crespin,levitz}
it has attracted interest\cite{choisnet,anisimov,lope}
because it is isostructural 
to CaCuO$_2$,\cite{siegrist} the parent ``infinite layer" material of high T$_c$ 
superconductors, and like CaCuO$_2$ has a formal
$d^9$ ion amongst closed ionic shells.
However, it is difficult to synthesize and
was not revisited experimentally until 
recently by Hayward $\it{et~al.}$ who produced it as the major phase by oxygen
deintercalation from LaNiO$_3$.\cite{hayward1}
Their materials consist of two phases, the majority being the 
infinite-layer (NiO$_2$-La-NiO$_2$) structure
and the minority being a disordered 
derivative phase.
Magnetization and neutron powder diffraction reveal no long-range 
magnetic order in their materials.
Its paramagnetic susceptibility has been fit by a Curie-Weiss form in the
$150 < T/K < 300$ range with S=$\frac{1}{2}$ and Weiss constant $\theta$ =
-257 K, but its low T behavior varies strongly from this form.
More recently, this same group has produced the isostructural and
isovalent nickelate NdNiO$_2$.\cite{hayward2}

One of the most striking features of LaNiO$_2$ is that it
potentially provides a structurally simple example of 
a {\it monovalent open shell transition metal d$^9$ ion}.
Except for the divalent Cu$^{2+}$ ion, the d$^9$ configuration is
practically nonexistent in ionic solids.
In particular, the formal similarity of Ni$^{1+}$ 
and Cu$^{2+}$ suggests that Ni$^{1+}$ compounds
might provide a ``platform" for additional high
temperature superconductors. It is these and related questions that we
address here.

In this paper we present results of theoretical studies of the electronic
and magnetic structures of LaNiO$_2$, and compare with the case of CaCuO$_2$
(or isovalent Ca$_{1-x}$Sr$_x$CuO$_2$)
which is well characterized.  A central question in transition metal
oxides is the role of correlation effects, which are certainly not known
{\it a priori} in LaNiO$_2$.  We look at results both from the local
density approximation (LDA) and its magnetic generalization, and then apply
also the LDA+U correlated electron band theory that accounts in a 
self-consistent mean-field way for Hubbard-like intraatomic repulsion
characterized by the strength U.  Our results 
reveal very different behavior between
LaNiO$_2$ and CaCuO$_2$, in spite of the structural and formal $d^9$ charge
similarities.  The differences can be traced to (1) the difference in
$3d$ site energy between Ni and Cu relative to that of Cu, 
(2) the ionic charge difference between
Ca$^{2+}$ and La$^{3+}$ and associated Madelung potential shifts, and (3) the 
participation of cation $5d$ states in LaNiO$_2$.

We also discuss briefly our discovery of anomalous 
behavior in the transition metal $3d^9$ ion as described
by LDA+U at large U.  Although well beyond the physical range of U for 
LaNiO$_2$, we find that LDA+U produces what might be characterized as
a $d^8$ ``singlet" ion in which the 
internal configuration is one $d(x^2-y^2)$ hole with spin up and one
$d(3z^2-r^2)$ hole with spin down, corresponding to an extreme spin-density
anisotropy on the transition metal ion but (nearly) vanishing net moment.

\section{Structure and calculation}

 In the samples of LaNiO$_2$ synthesized and reported by Hayward 
$\it{et}$ $\it{al.}$, there exist two phases with 
space group $P4/\it{mmm}$ (No. 123) but different site symmetry.\cite{hayward1} 
We focus on the majority infinite-layer phase, 
which is isostructural with CaCuO$_2$.\cite{siegrist} 
In the crystal structure shown in Fig. 1, 
Ni ions are at the corners of the square 
and La ions lie at the center of unit cell.
The bond length of Ni-O is 1.979 $\AA$, about 2\% more than 
that of Cu-O in CaCuO$_2$ (1.93 $\AA$). 
We used the lattice constants $a=3.87093{ }\AA,{ }
 c=3.3745\AA$,\cite{hayward1}
with a ($\sqrt{2}\times\sqrt{2}$) supercell 
space group $I4/\it{mmm}$ (No. 139) for AFM calculations.

\begin{figure}[bt]
\psfig{figure=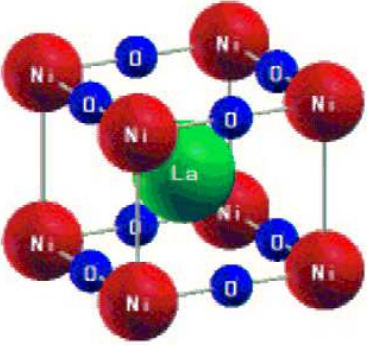,width=5.0cm,angle=0} 
\caption{\label{structure}
 (Color online)Crystal structure of LaNiO$_2$, isostructural to CaCuO$_2$. 
Ni ions are in the origin and La ions in the center of the unit cell.
It has no axial oxygens. }
\end{figure}
 
The calculations were carried out with the full-potential nonorthogonal 
local-orbital (FPLO) method\cite{klaus}
and a regular mesh containing 196 ${\bf k}$ points 
in the irreducible wedge of the Brillouin zone.
Valence orbitals for the basis set were La $3s3p3d4s4p4d5s5p6s6p5d4f$,
Ni $3s3p4s4p3d$, O $2s2p3s3p3d$. As frequently done 
when studying transition metal oxides, 
we have tried both of the popular forms of functional\cite{u1,u2}
of LDA+U method\cite{AZA} with 
a wide range of on-site Coulomb interaction U from 1 to 8 eV,
but the intra-atomic exchange integral J=1 eV was left unchanged.
For CaCuO$_2$, we used the same conditions 
as the previous calculation
done by Eschrig $\it{et}$ $\it{al}.$ using FPLO.\cite{eschrig}

\section{Results}
\subsection{LDA Results}
We present first the LDA results.
The paramagnetic (PM) band structure with its energy scale 
relative to Fermi energy $E_F$ is given in Fig. 2. 
A complex of La $4f$ bands is located at +2.5 eV 
with bandwidth less than 1 eV.  
The O $\it{2p}$ bands extend from about -8 eV to -3.2 eV.
The Ni $\it{3d}$ bands are distributed from -3 eV to 2 eV,
with the localized $t_{2g}$ complex near -1.5 eV, 
while the broad La $\it{5d}$
states range from -0.2 eV to 8 eV.
Unlike in PM CaCuO$_2$, there are two bands crossing E$_F$.
One is like the canonical $d(x^2-y^2)$ derived band in the cuprates,
rather broad due to the strong $\it{dp\sigma}$ antibonding interaction 
with oxygen $p_x, p_y$ states and enclosing holes centered at the M point.
The other band, lying at -0.2 eV at $\Gamma$ and also having its maximum
at the M=($\frac{\pi}{a},\frac{\pi}{a},0)$ point, 
is a mixture of La $\it{5d}(3z^2-r^2)$ states and some 
Ni $\it{3d}(3z^2 -r^2)$ character.  Already this band indicates importance
of Ni $3d$ - La $5d$ band mixing.

\begin{figure}[tbp]
\rotatebox{-90}{\resizebox{6cm}{7cm}{\includegraphics{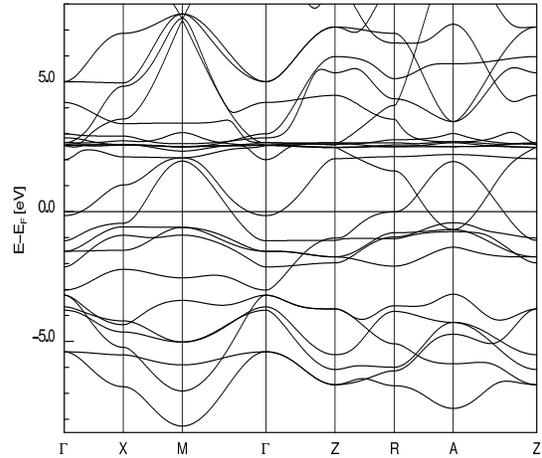}}}
\caption{LDA paramagnetic band structure of LaNiO$_2$.
The Ni $\it{3d}(x^2-y^2)$ band crosses the Fermi level (zero energy)
very much as occurs in cuprates (see Fig. 3).
The La $\it{4f}$ bands lie on $2.5-3.0$ eV. 
The La $5d(3z^2-r^2)$ band drops below E$_F$ 
at $\Gamma$ and A.}
\label{pmband}
\end{figure}

\begin{figure}[tbp]
\rotatebox{-90}{\resizebox{5cm}{7cm}{\includegraphics{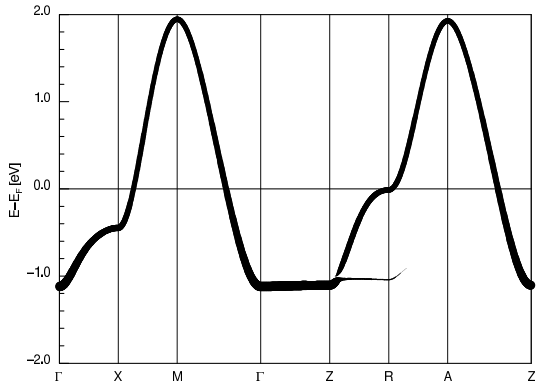}}}
\caption{``Fatband" representation of Ni $3d(x^2-y^2)$ in LDA. 
This band appears at first very two-dimensional,
but is not because (1) the saddle point at X($0,\pi/a,0$) 
is not midway between the $\Gamma$ and M($\pi/a,\pi/a,0$)
energies, and (2) $k_z$ dispersion between the X and R($0,\pi/a,\pi/c$).}
\label{pmdos}
\end{figure}

\begin{table}[bt]
\caption{Tight binding parameters (in meV) for Ni 3$d(x^2-y^2)$
of LaNiO$_2$ and Cu 3$d(x^2-y^2)$ of CaCuO$_2$.
$\varepsilon_{0}$ is the site energy and $t$'s are hopping integrals.
$\it{Ratio}$ (in $\%$) is hopping integrals for LaNiO$_2$ to those
for CaCuO$_2$.}
\begin{center}
\begin{tabular} {cccc}\hline\hline
 parameters~        &~ LaNiO$_2$  &~ CaCuO$_2$  &
             ~$|\it{Ratio}|$ ($\%$)~       \\ \hline
 $\varepsilon_{0}$ & 93 & -200&   \\
 $t(100)$          &381 & 534 & 71 \\
 $t(110)$          &-81 & -84 & 96 \\
 $t(001)$          &58  & 83  & 70 \\
 $t(101)$          & 0  & -2  & 0  \\
 $t(111)$          &-14 & -19 & 74 \\ \hline\hline
\end{tabular}
\end{center}
\label{table1}
\end{table}

Using a simple one-band tight binding model
\begin{eqnarray}
\varepsilon_k= \varepsilon_{\circ} -
  \sum_{R} t_R~ e^{i \vec k \cdot \vec R},\nonumber
\end{eqnarray}
the Ni $3d(x^2-y^2)$ band shown in Fig. 3 can be reproduced with a few
hopping amplitudes, but requiring more than might have been anticipated. 
The site energy is $\varepsilon_{\circ}=93$ meV, slightly above the Fermi 
level, and the hopping integrals (in meV)
are $t(100)=381$, $t(110)=-81$, $t(001)=58$ and 
$t(111)=-14$. There is no hopping along the
(101) direction.
As anticipated from the cuprates, the largest hopping is via $t(100)$.
However, to correctly describe the $k_z$ dispersion from X-R 
({\it i.e.} along $\pi/a,0,k_z$)
together with
the {\it lack of dispersion}  
from $\Gamma$ - Z ($0,0,k_z$) and also M-A ($\pi/a,\pi/a,k_z$),
the third neighbor hopping terms $t(111)$ must be included.

The comparison of the single band tight binding parameters with those of
CaCuO$_2$ is given in Table I.  It should be noted that the state in mind
is an $x^2 - y^2$ symmetry state that is orthogonal to those on 
neighboring Ni/Cu ions, {\i.e.} an $x^2 - y^2$ symmetry Wannier orbital.  In Ni,
the on-site energy is 0.3 eV above what it is in CaCuO$_2$, lying above
E$_F$ rather than below.  This difference is partially due to the different
Madelung potential in the two differently-charged compounds, but it also
reflects some intrinsic hole-doping in the nickelate that leads to a lower
Fermi level.  The largest
hopping amplitude (the conventional {\it t}) is 71\% of its value in the
cuprate, while the second ({\it t'}) is essentially the same.  The $t(001)
\equiv t_z$ is also 70\% of its value in the cuprate, while the other
amplitudes are the almost unchanged.  

\begin{figure}[tbp]
\resizebox{7.85cm}{4.85cm}{\includegraphics{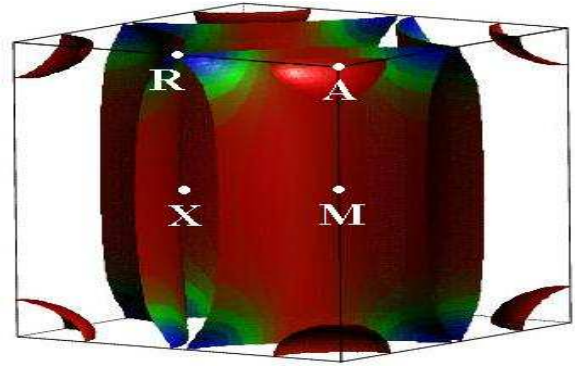}}
\caption{(Color online) Paramagnetic Fermi surface in the local
density approximation. 
In the center (not visible),
i.e. $\Gamma$, there is a sphere (a radius $0.25$($\pi/a$)) 
having $d(3z^2 -r^2)$ character of Ni and La. 
The cylinder with radius $0.8(\pi/a)$
contains Ni $d(x^2 -y^2)$ holes,
whereas another sphere (a radius $0.4$($\pi/a$))
at each corner contains Ni $d(zx)$ electrons.}
\label{FS}
\end{figure}

The LDA Fermi surfaces are shown in Fig. 4.  As for the cuprates, the
Fermi surface is dominated by the M-centered hole barrel.  In this system
neighboring barrels touch at R=($\pi/a,0,\pi/c$) because the saddle point at R 
happens to lie at E$_F$.  The Fermi surfaces also include two spheres 
containing electrons.
The sphere at $\Gamma$, with mixed Ni and La $d(3z^2 -r^2)$ character,
contains about $0.02$ electrons.
The A-centered sphere is mainly Ni $d(zx)$ in character and
contains approximately 0.07 electrons per Ni.
The barrel, whose radius of $0.8~\pi/a$ in the (1,1,$k_z$) direction is almost
independent of $k_z$ but which varies along (1,0,$k_z$),
possesses about 1.1 
holes, accounting for the total of the $1.0$ hole that is required by
Luttinger's theorem and also fits the 
formal Ni$^{1+}$ valence (which, being a metal and also mixing with La as well
as with O states, is not
very relevant).

To investigate magnetic tendencies, attempts to find both ferromagnetic (FM) and
antiferromagnetic (AFM) states were made.  A stable $\sqrt{2}\times\sqrt{2}$
AFM state was
obtained, with  
spin moment 0.53 $\mu_B$ per Ni.  This state
has lower energy by $6$ meV/Ni than that of PM state.
Just as for the paramagnetic case, the AFM state has 
entangled bands of La $\it{5d}$,
Ni $\it{3d}$ and O $\it{2p}$ character near the Fermi energy.
In contrast to the unpolarized case (and CaCuO$_2$), with AFM order the
large electron pocket has primarily La $5d(xy)$ character and the 
slightly occupied electron pocket at $\Gamma$ has a combination of
La $5d(3z^2-r^2)$ 
and Ni $3d(3z^2-r^2)$ character.
Attempts to obtain a FM solution   
always led to a vanishing moment.

\subsection{Consideration of Correlation with LDA+U}
As noted in the introduction, no magnetic order has been observed in
LaNiO$_2$, either by magnetization or by neutron scattering.  Although
the local density approximation often does quite well in predicting
magnetic moments,  for weakly or nearly magnetic systems renormalization by 
spin fluctuations becomes important\cite{moriya,mazin,mazin2}
and such effects are not included in
the local density approximation.  There is also the question of the
strength of correlation effects due to an intra-atomic repulsion $U$
on the Ni site.  Analogy to CaCuO$_2$ (same formal $d^9$ configuration,
neighboring ion in the periodic table), which is a strong antiferromagnetic
insulator, suggests that effects due to $U$ might have some importance.  Here we
apply the LDA+U ``correlated band theory'' method to assess effects of
intra-atomic repulsion and compare with observed behavior.
In the following subsection we compare and contrast with CaCuO$_2$.

\begin{figure}[tbp]
\rotatebox{-90}{\resizebox{6cm}{7cm}{\includegraphics{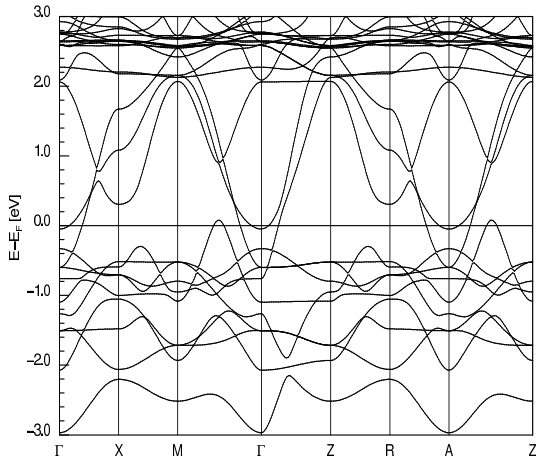}}}
\caption{LDA antiferromagnetic band structure of LaNiO$_2$.
The Ni $3d$ bands lie above -3 eV and are disjoint from the O $2p$ bands
(not shown) which begin just below -3 eV.  The antiferromagnetism introduces
the gap in the Ni $dp\sigma$ band midway between $\Gamma$ and M in the range
0 - 1 eV.
The symmetry points are given such as (0,0,x) for $\Gamma$(Z)
, (1/2,1/2,x) for X(R) and (1,0,x) for M(A). x is zero for the first
symbols and 1 for the symbols in parentheses.
}
\label{afmband}
\end{figure}

Upon increasing $U$ from zero in the antiferromagnetically ordered phase,
the spin magnetic moment of Ni increases from the LDA value of 0.53 $\mu_B$
to a maximum of
0.8 $\mu_{B}$ at $U=3$ eV. 
Surprisingly, for $U >$ 4 eV
the moment steadily decreases and by $U= 8$ eV it has {\it dropped} to 
0.2 $\mu_{B}$/Ni, which is less than half of its LDA value.
We emphasize that this behavior is unrelated to the observed
behavior of LaNiO$_2$ (which may need little or no additional correlation
beyond LDA).  However, this unprecedented response of the transition metal
ion to the
imposition of a large $U$ gives new insight into a feature of
the LDA+U method that has not been observed previously.

\begin{figure}[tbp]
\rotatebox{-90}{\resizebox{8.5cm}{7.5cm}{\includegraphics{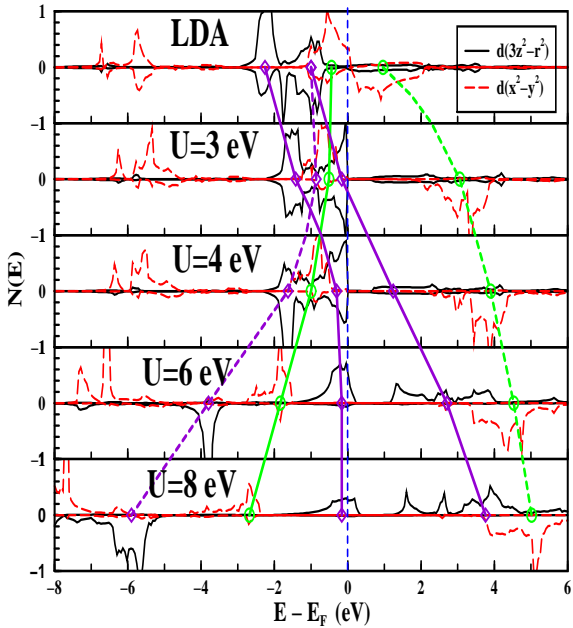}}}
\caption{(Color online)Change of the Ni $3d(3z^2-r^2)$ 
and  $3d(x^2-y^2)$ densities of states as on-site Coulomb 
interaction U increases.  One can easily identify a splitting (``Mott 
transition'') of the $3d(x^2-y^2)$ states occurring near $U$=0, and the
light (green) lines outline their path with increasing $U$ (majority is
solid, minority is dashed).  A
distinct Mott transition involving oppositely directed moment of the
$3d(3z^2-r^2)$ states is outlined with the dark (purple) lines. This 
moment is oppositely directed.  The conceptual picture is also complicated
by the splitting even at $U$=0 which persists in the majority states,
leaving a band at E$_F$ with strong Ni $3d(3z^2-r^2)$ character as well as
the expected upper Hubbard band at 4 eV.}
\label{2Mott}
\end{figure}

This ``quenching'' of the local moment with increasing $U$
results from 
behavior of Ni $3d(3z^2-r^2)$ states that is analogous to those of the
$3d(x^2-y^2)$, but with the direction of spin inverted (then with
additional complications). As usual for a $d^9$ ion in this environment, 
the majority $3d(x^2-y^2)$ state of Ni is fully occupied 
even at $U=2$ eV, while the minority state is completely unoccupied 
at $U=3$ eV, where the moment is maximum and the system is essentially
Ni$^{1+}$ S=$\frac{1}{2}$.  One can characterize this
situation as a Mott insulating $3d(x^2-y^2)$ orbital, as in the undoped
cuprates. 
At $U$ = 3 eV, the density of states
has a quasi one-dimensional van Hove singularity due to
a flat band just below (bordering) the Fermi energy as can be seen in 
the $3d$ DOS shown in Fig. \ref{2Mott}.  Upon increasing $U$ to 4 eV,
rather than reinforcing the $S=\frac{1}{2}$ configuration 
of Ni and thereby forcing
the La and O ions to cope with electron/hole doping, the Ni $d(3z^2-r^2)$ 
states begin to polarize.  The charge on the Ni ion drops somewhat, moving it in
the Ni$^{1+}$ $\rightarrow$ Ni$^{2+}$ direction, with the charge going into 
the La $5d$ -- O $2p$ states.
Idealizing a bit, one might characterize the movement of (unoccupied)
{\it majority} character of $3d(3z^2-r^2)$ well above E$_F$ as a Mott
transition of these orbitals, which is not only {\it distinct from}
that of the $3d(x^2-y^2)$ states, but is {\it oppositely} directed,
leading to an on-site ``singlet'' type of cancellation. 

\begin{figure}[tbp]
\rotatebox{-90}{\resizebox{6cm}{6cm}{\includegraphics{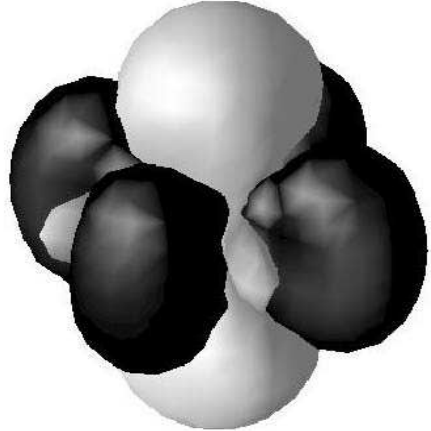}}}
\caption{Isocontour plot of the spin density of the
``singlet'' Ni ion ($U$ = 8 eV) when there is an $x^2-y^2$ hole with
spin up and a $3z^2-r^2$ hole with spin down.  Dark and light surfaces denote
isocontours of equal magnitude but opposite sign.
}
\label{spindens}
\end{figure}

This movement of states with increasing $U$ has been emphasized in
Fig. \ref{2Mott} for easier visualization.  The resulting spin density on the
transition metal ion at $U$ = 8 eV is pictured in 
Fig. \ref{spindens}.  There is strong 
polarization in all directions from the core except for the position of
nodes.  The polarization is strongly positive (majority) in the lobes
of the $3d(x^2-y^2)$ orbital, and just as strongly negative (minority spin)
in the lobes of the $3d(3z^2-r^2)$ orbital.  The net moment is (nearly) 
vanishing, but this results from a singlet combination (as nearly as it
can be represented within classical spin picture) of spin-half up in one
orbital and spin-half down in another orbital that violates Hund's first
rule.  The magnetization density
is large throughout the ion, but integrates to (nearly) zero.

This behavior is however more complicated than a Mott splitting of occupied
and unoccupied state, as can be seen from the 
substantial Ni $3d$ character that remains, even for $U$ = 8 eV, in a
band straddling E$_F$ while the rest of the weight moves to $\sim$4 eV.  
In both of these
bands there is strong mixing with La $5d(xy)$ states.  What happens is
that as the ``upper Hubbard $3d(3z^2-r^2)$ band'' rises as $U$ is increased,
it progressively mixes more strongly with the La $5d(xy)$ states, forming
a bonding band and an antibonding band.   While the antibonding combination
continues to move upward with increasing $U$,
the bonding combination forms a half-filled band which remains at E$_F$.  

Thus we have found that for the Ni$^{1+}$ ion in this environment,
increasing $U$ (well beyond what is physically plausible for LaNiO$_2$)
results in $S=\frac{1}{2}$
Ni$^{+1}$ being converted into a nominal Ni$^{+2}$ ion (the actual
charge changes little, however) in
which the two holes are coupled
into an intraatomic $S=0$ singlet.
This behavior involves yet a new kind of correlation
between the $\it{3d}(3z^2-r^2)$ states
and the $\it{3d}(x^2-y^2)$ states, but one which is due to
(driven by) the local environment.

This behavior is quite different from the results for $U$=8 eV reported
by Anisimov, Bukhvalov and Rice\cite{anisimov} using the Stuttgart
TBLMTO-47 code.  They obtained an AFM insulating solution analogous to
that obtained for CaCuO$_2$,\cite{eschrig} with a single hole in the $3d$
shell occupying the $3d(x^2-y^2)$ orbital that antibonds with the
neighboring oxygen $2p_{\sigma}$ orbital.  The reason for this different
result is not known, but it is now well established that multiple
solutions to the LDA+U equations often exist.\cite{shick1,shick2}

\section{Comparison with C\lowercase{a}C\lowercase{u}O$_2$ and Discussion}

\begin{figure}[tbp]
\rotatebox{-90}{\resizebox{9cm}{6cm}{\includegraphics{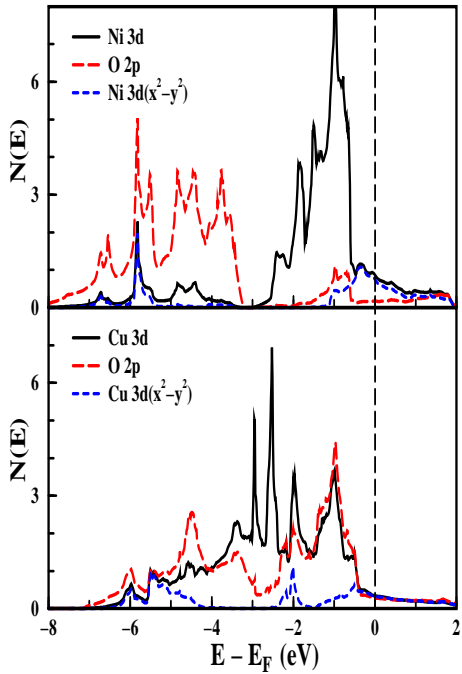}}}
\caption{(Color online)Comparison of LDA projected paramagnetic DOS LaNiO$_2$
(upper panel) and CaCuO$_2$ (lower panel).  Note the separation of the Ni
$3d$ states from the O $2p$ states in the upper panel, which does not occur
for the more strongly hybridized cuprate.}
\label{pjdos}
\end{figure}

Although Ni$^{+1}$ is isoelectronic to Cu$^{+2}$,
both the observed and the calculated behavior 
of LaNiO$_2$ are very different from
CaCuO$_2$.
In contrast to CaCuO$_2$, LaNiO$_2$ is (apparently) metallic, with
no experimental evidence of magnetic ordering for LaNiO$_2$. 
The differing electronic and magnetic properties mainly arise from two
factors.
First, the Ca $\it{3d}$ bands lying in the range of 4 eV and 9 eV
are very differently distributed from the
broader and lower La $\it{5d}$ bands in the range of -0.2 eV and 8 eV.
Secondly, in CaCuO$_2$, O $\it{2p}$ states extend to
Fermi level and overlap strongly with Cu $\it{3d}$ states, and
the difference of the two centers is less than 1 eV, as can be seen
in Fig. \ref{pjdos}.
Thus, there is a strong $2p-3d$ hybridization that has been heavily
discussed in high T$_c$ materials.
In LaNiO$_2$, however, Ni $\it{3d}$ states lie just below
the Fermi level, with O $\it{2p}$ states located 
$3-4$ eV below the center of Ni bands.
Therefore, p-d hybridization, which plays a crucial role in the 
electronic structure and superconductivity of CaCuO$_2$, becomes much weaker.

\section{Summary}
Aside from the formal similarity to CaCuO$_2$, the interest in LaNiO$_2$
lies in the occurrence of the unusual 
monovalent Ni ion.  As we have found and in
apparent agreement with experiment, this compound is a metal, and the 
``charge state'' of a transition metal atom in a metal usually has much
less significance than it is in an insulator.  
It may be because the compound is
metallic that it is stable, but in this study we are not addressing 
energetics and stability questions.

Hayward $\it{et}$ $\it{al}.$\cite{hayward1}
had already suggested that the experimental findings
could arise from reduced covalency between the Ni $3d$ and O $2p$ orbitals,
and the 30\% smaller value of the hopping amplitude $t$ indeed reflects
the smaller covalency, as does the increased separation between the Ni $3d$
and O $2p$ bands.
It is something of an enigma that in CaCuO$_2$ and other 
cuprates, LDA calculations
fail to give the observed antiferromagnetic states, while in LaNiO$_2$ LDA
predicts a weak antiferromagnetic state when there is no magnetism observed.
In the cuprates the cause is known and is treated in a reasonable
way by application of the LDA+U
method.  In this nickelate, application of the LDA+U method does not seem to
be warranted (although novel behavior occurs it if it used).
Rather, the prediction of weak magnetism adds this compound to the small but
growing number of systems (ZrZn$_2$,\cite{zrzn2} Sc$_3$In,\cite{sc3in} and 
Ni$_3$Ga,\cite{mazin2} for example)
in which the tendency toward magnetism is
overestimated by the local density approximation.  It appears that this
tendency can be corrected by accounting for magnetic 
fluctuations.\cite{moriya,mazin2}

\section{Acknowledgment}
We acknowledge useful communication with M. Hayward during the course of
this research, and discussions with J. Kune\v{s} and P. Novak about the 
behavior of the LDA+U method.
This work was supported by National Science
Foundation Grant DMR-0114818.

\end{document}